\documentclass[epj]{svjour}
\usepackage{graphics}
\newcommand{\sech}{\mbox{sech}}
\newcommand{\sub}[1]{_{\scriptscriptstyle{#1}}}
\renewcommand{\sup}[1]{^{\:\:\scriptscriptstyle{#1}}}
\renewcommand{\sup}[1]{^{\scriptscriptstyle{#1}}}
\newcommand{\lc}[1]{\mathcal{L}^{\scriptscriptstyle 2}(#1)}
\usepackage{graphicx}
\usepackage{dcolumn}
\usepackage{bm}
\usepackage{epsfig}
\usepackage{graphics}
\usepackage{amsfonts}
\begin{document}
\title{Particle resonance in the Dirac equation in the presence of a delta interaction and 
a perturbative hyperbolic
potential}
\titlerunning{Particle resonance in the Dirac equation...}
\author{V\'{\i}ctor M Villalba\inst{1} \inst{2}\thanks{\email{villalba@ivic.ve}} \and Luis A. Gonz\'alez-D\'{\i}az\inst{2}
}                     
%
%
\institute{Department of Physics and Astronomy, University of Waterloo,
Waterloo, Ontario Canada N2L 3G1 \and Centro de F\'{\i}sica IVIC Apdo 21827, Caracas 1020A,
Venezuela}
\date{Received: date / Revised version: date}
%
\abstract{
In the present article we show that the energy spectrum of the
one-dimensional Dirac equation, in the presence of an attractive
vectorial delta potential, exhibits a resonant behavior when one
includes an asymptotically spatially vanishing weak electric field
associated with a hyperbolic tangent potential. We solve the Dirac
equation in terms of Gauss hyper-geometric functions and show
explicitly how the resonant behavior depends on the strength of the
electric field evaluated at the support of the point interaction. We
derive an approximate expression for the value of the resonances
and compare the results calculated for the hyperbolic potential with those 
obtained for a linear perturbative potential. 
  Finally, we characterize the resonances with the help of
the phase shift and the Wigner delay time.
\PACS{
      {03.65.Pm}{Relativistic wave equations }   \and
      {03.65.Ge}{Solutions of wave equations: bound states}
     } 
} 
\maketitle
\section{Introduction}
\label{intro}
Spontaneous particle production in the presence of strong
electromagnetic fields is undoubtedly one of the most interesting
phenomena associated with the quantum vacuum
\cite{Greiner}. The study quantum effects associated
with relativistic particles in strong fields dates back to the
pioneering work of Klein \cite{Klein} where he studied the
reflection and penetration of electron waves on a potential barrier,
obtaining the result that, for very strong potentials a large number of
electrons with negative energy penetrate into the wall. This effect
is called the  Klein paradox.  Sauter \cite{Sauter} found
essentially the same results in the most general case when the  barrier
is smoothed. Particle creation by strong infinitely extended
constant electric field, which is closely related to the Klein
paradox, was predicted by Heisenberg and Euler \cite{Heisenberg}.

The study of supercritical effects and resonant particle production
by strong Coulomb-like potentials  dates back to the pioneering
works of Pieper and Greiner \cite{Pieper} and Gershtein and
Zeldovich \cite{Gershtein} where it was shown that spontaneous
positron production was possible when two heavy bare nuclei with
total charge larger than some critical value $Z_{c}$ collided with
each other. The critical $Z_{c}=1/\alpha \approx 137$ is the value for which the $1S$
state of the hydrogenlike  atom with potential $V=-Ze/r$ has energy
$E=-m c^2$. Supercriticality effects are based on spontaneous positron
emission induced by the presence of very strong attractive electric
potentials. The energy level of an unoccupied bound state sinks into
the negative energy continuum. An electron of the Dirac sea is
trapped by the potential, leaving a positron that escapes to
infinity.  The electric field responsible for supercritical effects has a strength 
of $E \sim m_{e}^2 c^{3}/(e \hbar) $ and,  in the region where the field acts, 
it makes a work  larger than $2m_{e}c^{2}$  which is the value of the gap
between the negative and positive energy continua.  Such strong
electric fields could be produced in super-heavy nuclei, heavy-ion
collisions \cite{Greiner,Greiner2} and in astrophysical phenomena. 

In order to get a deeper understanding of the mechanism responsible
for resonant peaks appearing in the energy spectrum when
supercritical fields are present, and the role played by
perturbative fields in the shape of resonances, we proceed to work
with a vector point interaction in the presence of a homogeneous,
asymptotically vanishing electric field \cite{villalba,Cannata,Jiang} Point interactions
potentials permit us  to tackle, in a simple way, more complex
short-ranged potentials. Among the advantages of working with
confining delta vector potentials we have the fact that, they only possess a
single bound state and the treatment of the interaction reduces to a
boundary condition. Bound states of the relativistic wave equation in
the presence of point interactions have been carefully discussed in
the literature \cite{Sutherland,McKellar,McKellar2,FD:89}. The
one-dimensional Dirac equation in the presence of a vector point
delta interaction has also been a subject of study in the search for
supercritical effects induced by attractive potentials. In this
case, we see that a vectorial delta potential is strong enough to
pull the bound state into the negative energy continuum $E=-m c^2$
\cite{FD:89,Calkin,Nogami}, nevertheless this supercritical state
does not evolve to a real resonant state. Since we are interested in
studying the mechanism of positron production by supercritical
fields, we proceed to analyze the resonant behavior of the energy
when a bound state dives into the negative continuum. This resonant
behavior is associated with the appearance of simple poles of the
resolvent on the unphysical sheet at a position very near the real
axis \cite{Galindo}, which can be identified as positron states with a
short mean life.

In the present article, applying  the idea developed by Titchmarsch
\cite{ET:58} and Barut \cite{Barut}, we compute the energy spectrum
of the one-dimensional Dirac equation in the presence of a vector
Dirac delta interaction and a weak electric field associated with a linear 
and an hyperbolic potential.  We find that,  in both  cases the energy spectrum
exhibits a resonant behavior. This problem can be considered the relativistic 
extension of the one-dimensional Stark effect \cite{Kleber,Calvancanti,Dunne}. 

The paper is structured as follows: In section 2, we solve the
one-dimensional Dirac equation in the presence of an attractive
$\delta$ potential and a hyperbolic tangent vector potential. In
section 3, we compute the energy resonances and show how they depend
on the electric field strength evaluated at the delta interaction
point. We also derive an approximate analytic expression for the
energy resonances. In section 4, we characterize the resonance by
computing the phase shift $\phi$ and the Wigner time delay 
for the hyperbolic potential. 
Finally, in section 5 we summarize our conclusions.

\section{The one-dimensional Dirac equation}

The energy spectrum of the Dirac equation in the present of a delta
potential has been studied by different authors in the literature.
The attractive vectorial delta interaction  is only is able to
support a bound state which becomes a zero-momentum resonance for
$E=-m$ but it is not strong enough to sink the bound state into the
Dirac see to create a time-decaying resonant state. In this section
we proceed to solve the 1+1 Dirac equation in the presence of an
attractive vector point delta interaction and a potential $V(x)$ which 
we will consider perturbative. The Dirac equation, expressed in natural
units ($\hbar=c=1$) takes the form \cite{Greiner1}
\begin{equation}
\left(  i\gamma^{\mu}(\frac{\partial}{\partial
x^{\mu}}-ieA_{\mu})-m\right)
\Psi=0,\label{1a}%
\end{equation}
where $A_{\mu}$ is the vector potential that in our case takes the form 
$eA^{\mu}=(-g \delta(x)+V(x))\delta^{\mu}_{0}$, $e$ is the charge, and $m$
is the mass of the electron. The Dirac matrices $\gamma^{\mu}$
satisfy the commutation relation $\left\{
\gamma^{\mu},\gamma^{\nu}\right\}  =2\eta^{\mu\nu}$ with
$\eta^{\mu\nu }=\mathrm{diag}(1,-1).$ Since we are working in 1+1
dimensions, we choose to work in a two-dimensional representation of
the Dirac matrices
\begin{equation}
\gamma^{0}=\sigma_{3},\gamma^{1}=-i\sigma_{2}.\label{2}%
\end{equation}
Substituting the matrix representation (\ref{2}) into equation
(\ref{1a}), and taking into account that the potential interaction
does not
depend on time, we obtain%

\begin{equation}
\{-i\sigma_{{1}}\frac{d}{dx}+\left(  V(x)\ -E\right)  +m\sigma_{{3}%
}\}\mathsf{X}(x)=0,\label{diraca}%
\end{equation}
with $\Psi=\sigma_{3}\mathsf{X}$, and
\begin{equation}
\label{equis} \mathsf{X}(x)= \left(
\begin{array}{cc}
X_{{1}}\\
X_{{2}}
\end{array}
\right),
\end{equation}
where E is the energy eigenvalue.  The components of  Eq. (\ref{equis}) satisfy the boundary conditions:
\begin{eqnarray}
\label{dominio}
X_{{1}}(0^{{+}}) &  =X_{{1}}(0^{{-}})\cos{g}-iX_{{2}}(0^{{-}}%
)\hspace{0.1cm}\mathrm{\sin}{g},\\ \nonumber
X_{{2}}(0^{{+}}) &  =-iX_{{1}}(0^{{-}})\hspace{0.1cm}\mathrm{\sin}{g}
+X_{{2}}(0^{{-}})\cos{g}.
\end{eqnarray}
The relations given by Eq. (\ref{dominio}) describe an attractive delta
vector potential interaction of strength $g$ \cite{FD:89}.

Taking into account
the gamma matrices representation (\ref{2}),  we obtain the system of equations
\begin{equation}
\left(  m+V(x)-E\right)  X_{{1}}-i\frac{dX_{2}}{dx}%
=0,\label{pdiraca}%
\end{equation}
\begin{equation}
i\frac{dX_{1}}{dx}+\left(  m-V(x)+E\right)  X_{{2}}%
=0,\label{sdiraca}%
\end{equation}
Introducing the functions $\Omega_{{1}}$ and $\Omega_{{2}}$
\begin{equation}
X_{{1}}=\Omega_{{1}}+i\Omega_{{2}},\quad X_{{2}}=\Omega_{{1}}-i\Omega_{{2}%
},\label{rela}%
\end{equation}
we see that the system of equations
given by Eqs. (\ref{pdiraca})-(\ref{sdiraca}) reduces to
\begin{eqnarray}
\frac{d\Omega_{1}}{dx}+i\left(  V(x) -E\right)  \Omega_{{1}%
}-m\Omega_{{2}} &  =0,\label{podiraca}\\
\frac{d\Omega_{{2}}}{dx}-i\left(  V(x) -E\right)  \Omega_{{2}%
}-m\Omega_{{1}} &  =0,\label{sodiraca}%
\end{eqnarray}
which is more tractable in the search of exact solutions.
Using  Eq. (\ref{podiraca}) and  Eq. (\ref{sodiraca}) we
obtain that  $\Omega\sub{1}(x)$ satisfies the second order differential equation
\begin{eqnarray}\label{omega1-lambda-x}
\frac{d\sup{2}}{dx\sup{2}}\Omega\sub{1}(x)+\{i \frac{dV(x)}{dx}+\left(V(x)-E\right)\sup{2}-m\sup{2}\}\,\Omega\sub{1}(x)=0
\end{eqnarray}

\subsection{Delta interaction and linear potential $\lambda x$}

We are interested in solving the Dirac equation (\ref{diraca}) in
the presence of an attractive delta and a linear perturbative
potential $V(x)=\lambda\ x$,  which corresponds to a constant
electric field $eE_{x}=-\lambda$. The perturbation of the energy
spectrum of a non-relativistic particle by a constant electric field
has been extensively discussed in the literature, mainly because to
its relation to the Stark effect and the presence of energy
resonances \cite{Galindo}.

The solution to Eq. (\ref{omega1-lambda-x}), for $\lambda<0$,
satisfying the boundary condition of a growing oscillatory solution
for $x<0$ and a damping solution for $x>0$ can be expressed in terms
of the parabolic cylinder functions $D_{\nu}(x)$ \cite{Abramowitz}
as
\begin{eqnarray}\label{sol1}
\Omega\sup{+}\sub{1}(x)=A\,D\sub{\rho}\left({\sqrt{\frac{2}{\lambda}}}e\sup{\frac{i\pi}{4}}\left(
\lambda\,x-E\right)\right),\quad x\geq0. \nonumber \\ 
\Omega\sup{-}\sub{1}(x)=B\,D\sub{\rho}\left(-{\sqrt{\frac{2}{\lambda}}}e\sup{\frac{i\pi}{4}}\left(
\lambda\,x-E\right)\right),\quad x\leq0.
\end{eqnarray}
where  $D\sub{\rho}$ are the parabolic cylinder functions
\cite{Abramowitz}, $\rho=\frac{im\sup{2}}{2\lambda}$, $A$ and  $B$
are constants to be determined using  the boundary condition given
by Eq. (\ref{dominio}). The solution $\Omega\sup{+}\sub{1}(x)$
belongs to  $\lc{0,\infty}$ with  $Im\, E>0$. The solution
$\Omega\sup{-}\sub{1}(x)$ behaves asymptotically as an outgoing
wave, a condition that defines a  Siegert state \cite{Siegert}

\begin{figure}[htpb]
\vspace{0.2cm}
\begin{center}
\includegraphics[width=8cm]{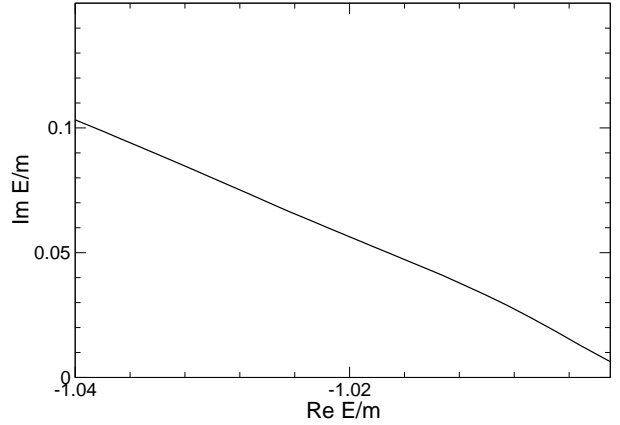}
\end{center}
\caption{ Behavior of $\Im(E)$ against $\Re(E)$ for the resonant energy states of a delta interaction with $g=-0.99 \pi$ perturbed by a 
linear potential $\lambda x$ with $0<\lambda<0.03$. $\Im(E)$ and $-\Re(E)$ increase as the parameter $\lambda$ increases} \label{Fig1a}
\end{figure}

Inserting  Eq. (\ref{sol1}) into  Eq. (\ref{podiraca}) and using the
recurrence relations for the parabolic cylinder functions
\cite{Abramowitz}, we obtain
\begin{eqnarray}\label{sol2}
\Omega\sup{+}\sub{2}(x)=A \frac{m e^{i \pi/4}}{\sqrt{2
\lambda}}D_{\rho
-1}\left({\sqrt{\frac{2}{\lambda}}}e\sup{i\frac{i\pi}{4}}\left(
\lambda\,x-E\right)\right) \\
\Omega\sup{-}\sub{2}(x)=B\ \frac{m e^{-i \pi/4}}{\sqrt{2
\lambda}}D_{\rho -1}\left(-{\sqrt{\frac{2}{\lambda}}}e\sup{i\frac{i
\pi}{4}}\left( \lambda\,x-E\right)\right)
\end{eqnarray}
Using the boundary condition (\ref{dominio})  we obtain that the energy resonances
satisfy the equation
\begin{eqnarray}
\label{eigenv}
(X_{1}^{-}(0)X_{2}^{+}(0)-X_{2}^{-}(0)X_{1}^{+}(0))\cos(g) \nonumber \\
+i(X_{1}^{-}(0)X_{1}^{+}(0)-X_{2}^{-}(0)X_{2}^{+}(0))\sin(g)=0,
\end{eqnarray}
and with the help of the definition of $X_{1}$, and $X_{2}$ in terms of $\Omega_{1}$ and $\Omega_{2}$  (\ref{rela}), 
we obtain the result  that the eigenvalue equation (\ref{eigenv}) can be written as: 
\begin{equation}\label{autovalores}
\Omega\sup{-}\sub{2}(0)\Omega\sup{+}\sub{1}(0)-e^{-2ig}\Omega\sup{-}\sub{1}(0)\Omega\sup{+}\sub{2}(0)=0
\end{equation}
which, after substituting the expressions for $\Omega_{1}$ and
$\Omega_{2}$ in terms of the parabolic cylinder functions reduces to
\begin{eqnarray}\label{autoval}
\lefteqn{D\sub{\rho}\left({-\sqrt{\frac{2}{\lambda}}}e\sup{i\frac{1\pi}{4}}
E\right)
D\sub{\rho-1}\left({\sqrt{\frac{2}{\lambda}}}e\sup{i\frac{1\pi}{4}}E\right)} \nonumber \\
& -i e^{-2ig}
D\sub{\rho}\left({\sqrt{\frac{2}{\lambda}}}e\sup{i\frac{1\pi}{4}}E\right)
D_{\rho
-1}\left({-\sqrt{\frac{2}{\lambda}}}e\sup{i\frac{i\pi}{4}}E\right)=0
\end{eqnarray}
The eigenvalue equation (\ref{autoval})  permits us to obtain the
energy solutions to the Dirac equation (\ref{diraca}) in the
presence of a vectorial delta potential and a linear potential
$\lambda x$. For small values of $\lambda$ the linear potential
plays the role of perturbing the Dirac hamiltonian with a vectorial
delta interaction $-g \delta(x)$ creating resonant states that
are supercritical when $g$ approaches $\pi$ and $E\rightarrow
-m$, The solutions to Eq. (\ref{autoval}) for small values of
$\lambda$ should be obtained numerically.  Fig.  \ref{Fig1a} shows $\Im{E}$
versus $\Re{E}$ for varying $\lambda$ showing a resonant behavior analogous to
the one expected for the supercritical hydrogen atom. 

\subsection{Delta interaction and a hyperbolic potential}

Now we proceed to solve the 1+1 Dirac equation in the presence of
an
attractive vector point interaction potential represented by $eV(x)=-g%
\delta(x),$ and an asymptotically vanishing electric field
associated with the potential
\begin{equation}
\label{hiper} V(x)=\Lambda\tanh(kx).
\end{equation}
The potential (\ref{hiper}) is asymptotically constant for large
values of $x$ and has the linear potential as a limit for small
values of $k$.

After substituting the potential (\ref{hiper}) into Eq.
(\ref{omega1-lambda-x}),  we obtain the second order differential equation

\begin{equation}\label{ddodirac1}
\frac{d^{2}\Omega_{1}(x)}{dx^{2}}+\{ik\Lambda
\mathrm{sech}^{2}(kx)+(\Lambda\tanh(kx)-E)^{2}-m^{2}\}\Omega_{1}(x)=0
\end{equation}
In order to be able to identify the solutions of (\ref{ddodirac1})
exhibiting a damping  asymptotic behavior as
$x\rightarrow +\infty$, we introduce, for $x>0$, the new variable
$z=-e^{-2kx}$.  We see that
$\Omega_{1}(z)$ satisfies the second order  differential equation
\begin{eqnarray}\label{ddodirac1nueva}
\lefteqn{4k^{2}z^{2}\frac{d^{2}\Omega_{1}(z)}{dz^{2}}+4k^{2}z\frac{d\Omega_{1}(z)}{dz}+} \nonumber   \\ 
& \left[-i\frac{4z\Lambda
k}{(1-z)^{2}}+\left(\Lambda\frac{1+z}{1-z}-E\right)^{2}-m^{2}\right]\Omega_{1}(z)=0.
\end{eqnarray}
For $k>0$ and $\Lambda<0$, the solution of
(\ref{ddodirac1nueva}), vanishing as $x\rightarrow\infty$, with
$\Re{E}\leq-m$ and $\Im{E}\geq 0$, can be expressed in terms of
Gauss  hyper-geometric functions
 $_{2}F_{1}(a,b,c,z)$ \cite{Abramowitz} as
\begin{equation}\label{solomegader}
\Omega_{1}^{+}(z)=C_{1}(z-1)^{-\frac{i\Lambda}{k}}z^{a}{}_2F_1\left(c,
d,f;z\right),
\end{equation}
where
\begin{eqnarray}
a&=&\frac{\sqrt{m^{2}-(E-\Lambda)^{2}}}{2k}, \
b=\frac{\sqrt{m^{2}-(E+\Lambda)^{2}}}{2k},\nonumber \\
c&=&a-b-\frac{i\Lambda}{k},\ 
d=a+b(E,\Lambda,m,k)-i\frac{\Lambda}{k},\nonumber \\
f&=&1+2a.
\end{eqnarray}
Using Eq. (\ref{podiraca}) we can obtain the expression for
$\Omega_{2}^{+}(z)$
In order to solve Eq. (\ref{ddodirac1}) for $x<0$,  we introduce the new variable $\widetilde{z}=-e^{2kx}$, obtaining the
following differential equation
\begin{eqnarray}\label{eqder}
\lefteqn{4k^{2}\widetilde{z}\,^{2}\frac{d^{2}\Omega_{1}(\widetilde{z})}{d\widetilde{z}^{2}}+  
4k^{2}\widetilde{z}\,\frac{d\Omega_{1}(\widetilde{z})}{d\widetilde{z}}} \nonumber \\
&+\left[-i\frac{4\widetilde{z}\Lambda
k}{(\widetilde{z}-1)^{2}}+\left(\Lambda\frac{1+\widetilde{z}}{\widetilde{z}-1}-E\right)^{2}-m^{2}\right]
\Omega_{1}(\widetilde{z})=0
\end{eqnarray}
whose solution, exhibiting   the behavior of  an increasing reflected wave  as
$x\rightarrow-\infty$, with $\Re{E}\leq-m$ and $\Im{E}\geq 0$, is
\begin{equation}\label{solomegaizq}
\Omega_{1}^{-}(\widetilde{z})=C_{2}(\widetilde{z}-1)^{-\frac{i\Lambda}{k}}\widetilde{z}\,^{-b}{}_2F_1\left(
g,h,p;\widetilde{z}\right),
\end{equation}
with
\begin{eqnarray}
g&=&-b+a-\frac{i\Lambda}{k},\
h=-b-a-\frac{i\Lambda}{k},\nonumber \\
p&=&-1-2b.
\end{eqnarray}
The energy eigenvalues equation corresponding to the Dirac equation
Eq. (\ref{1a}), with the boundary conditions 
(\ref{dominio}) can be obtained after substituting $\Omega_{1}$ and
$\Omega_{2}$ into Eq. (\ref{autovalores}). When the potential $V(x)$
vanishes at $x=0$, such as in the linear potential or hyperbolic cases, 
the eigenvalue equation
(\ref{autovalores}) can be written in terms of
$\Omega\sup{-}\sub{1}(x)$ and $\Omega\sup{+}\sub{1}(x)$ as
\begin{eqnarray}
\label{auto}
\lefteqn{\frac{d\Omega\sup{-}\sub{1}(x)}{dx}\Omega\sup{+}\sub{1}(x)-e^{-2ig}\frac{d\Omega\sup{+}\sub{1}(x)}{dx}\Omega\sup{-}\sub{1}(x)} \nonumber \\ 
& -iE(1-e^{-2ig})\Omega\sup{-}\sub{1}(x)\Omega\sup{+}\sub{1}(x)=0
\end{eqnarray}

\section{Approximate solutions}

It is not straightforward to obtain an approximate expression for
the energy eigenvalues of Eq. (\ref{autoval}) in the vicinity of
$E=-m$ because the expansion parameter $\lambda$ appears in Eq.
(\ref{autoval}) in the argument and order of the parabolic cylinder
functions. It is also not  possible to try to apply perturbation
theory to find the complex energy eigenvalues. In this section, 
we derive an approximate solution to Eq.(\ref{diraca}) in terms 
of Airy functions for the linear potential
$\lambda\ x$ and for the hyperbolic potential $\Lambda \tanh (k x)$
for small $\Lambda$ and $k$.

Since we are interested in studying  Eq. (\ref{omega1-lambda-x}) for
very small values of $\lambda$ and search solutions  of Eq.
(\ref{autovalores}) with respect to the resonant energy value
$E=-m$, for the linear potential $V(x)=\lambda x$,  we approximate in 
Eq.  (\ref{omega1-lambda-x}), 
$\left(\lambda\,x-E\right)\sup{2}$  by
$-2\,\lambda\,E\,x+E^2$, obtaining in this way the
approximate differential equation
\begin{eqnarray}\label{omega1-lambda-x-aproximada}
\frac{d\sup{2}}{dx\sup{2}}{\Omega}\sub{1}(x)+\{i\lambda 
-2 \lambda\, E\,x+E^2-m\sup{2}\}\,{\Omega}\sub{1}(x)=0.
\end{eqnarray}
We proceed to solve Eq.
(\ref{omega1-lambda-x-aproximada}) demanding the solutions to
satisfy the resonance asymptotic conditions, that is, we choose
damping solutions for $x>0$ and diverging oscillating functions for
$x<0$. Those solutions are:
\begin{equation}
\label{sol1-aproximada}
{\Omega}\,\sup{+}\sub{1}(x)=\mathcal{A}\,Ai\left[-\frac{i\lambda+E\sup{2}-m\sup{2}-2\,E\,\lambda\,x}
{\left(2\,E\,\lambda\right)\sup{2/3}}\right]\\
\end{equation}
\begin{equation}
\label{sol2-aproximada} 
{\Omega}\,\sup{-}\sub{1}(x)=\mathcal{B}Ci^{+}\left[-\frac{i\lambda+E\sup{2}-m\sup{2}-2\,E\,\lambda\,x}
{\left(2\,E\,\lambda\right)\sup{2/3}}\right]
\end{equation}
where $\mathcal{A}$ and $\mathcal{B}$ are constants and $Ai(x)$ and
$Ci^{+}(x)=Bi(x)+iAi(x)$ are the Airy functions \cite{Abramowitz}.

Using  Eq. (\ref{sol1-aproximada}) and Eq.  (\ref{sol2-aproximada}) and
the eigenvalue equation (\ref{auto}), we obtain that the approximate
spectral equation takes the form

\begin{eqnarray}\label{ec-espectral-aproximada}
\lefteqn{\frac{\left(2\,E\,\lambda\right)\sup{1/3}\,i\,e\sup{-i\,g}}{\pi}+} \\ 
& 2i\,Ai\left(\Xi\right)\left(i\left(2\,E\,\lambda\right)\sup{1/3}\,Ci^{+'}\left(\Xi\right)+E Ci^{+}\left(\Xi\right)\right)\sin\,g=0 \nonumber
\end{eqnarray}
where  $\Xi=\frac{m\sup{2}-E\sup{2}+i\lambda}
{\left(2\,E\,\lambda\right)\sup{2/3}}$.

For the hyperbolic potential $\Lambda \tan(kx)$ we also have that 
since the parameters  $\Lambda$ and  $k$ are small, we can
approximate the expression $(\Lambda \tanh (kx)-E)^2$ by
$-2\Lambda k xE+E^2$ in Eq. (\ref{ddodirac1}),  obtaining in
this way the approximate differential equation for $\Omega_{1}$,

\begin{eqnarray}\label{omega1-tanh-x-aproximada}
\frac{d\sup{2}}{dx\sup{2}}{\Omega}\sub{1}(x)+\{i\Lambda\,k-2\Lambda\,k\,E
x+E^2-m\sup{2}\}\,{\Omega}\sub{1}(x)=0
\end{eqnarray}
which , after making the identification $\Lambda k
\rightarrow \lambda$ reduces to Eq. (\ref{omega1-lambda-x-aproximada}), therefore  the approximate eigenvalue
equation is given by Eq.  (\ref{ec-espectral-aproximada}) with $\lambda
= \Lambda k$.

In order to show the accuracy of the approximations
(\ref{omega1-tanh-x-aproximada}) and (\ref{omega1-lambda-x-aproximada}), 
we can compare the solutions
(\ref{sol1-aproximada}) and (\ref{sol2-aproximada}) with the exact
solutions given by Eq. (\ref{sol1}). Fig. 2 shows that, for small
values of $\lambda$, the solutions (\ref{sol1-aproximada})
and  (\ref{sol2-aproximada}) are a good  approximation to the parabolic cylinder functions
(\ref{sol1}). and the Gauss functions (\ref{solomegader}) and (\ref{solomegaizq}) for small $\Lambda$ and $k$.
 A comparison between the resonant energy values obtained after solving 
Eq. (\ref{ec-espectral-aproximada}) and Eq. (\ref{autoval}), shows that, 
for  $\lambda=-0.01$,  the eigenvalue equation (\ref{ec-espectral-aproximada}) 
gives resonant energies  with an error smaller than  $0.1\%$ .  For $\lambda=-0.01$ 
the solution to Eq. (\ref{autoval}) gives $E=-1.016964+0.049316 I$ and the approximate solution
obtained after solving Eq. (\ref{ec-espectral-aproximada}) is $E=-1.016139+0.049965 I$.  
Eq. (\ref{ec-espectral-aproximada})  gives more accurate energy eigenvalues as $\lambda \rightarrow 0$.
The advantage of using  Eq. (\ref{ec-espectral-aproximada}) for calculating the energy eigenvalues for small
values of $\lambda$, instead of solving Eq.  (\ref{autoval}),   lies in the fact that, for large values of the index $\rho$, 
the parabolic cylinder functions exhibit a 
divergent behavior  that  makes it  troublesome to obtain accurate values for the  
energy spectrum.   For values of $|\lambda|<0.01$ we made use of the approximate
solutions in order calculate the resonance energies. 

\begin{figure}[htbp]
\begin{center}$
\begin{array}{cc}
\includegraphics[width=4cm]{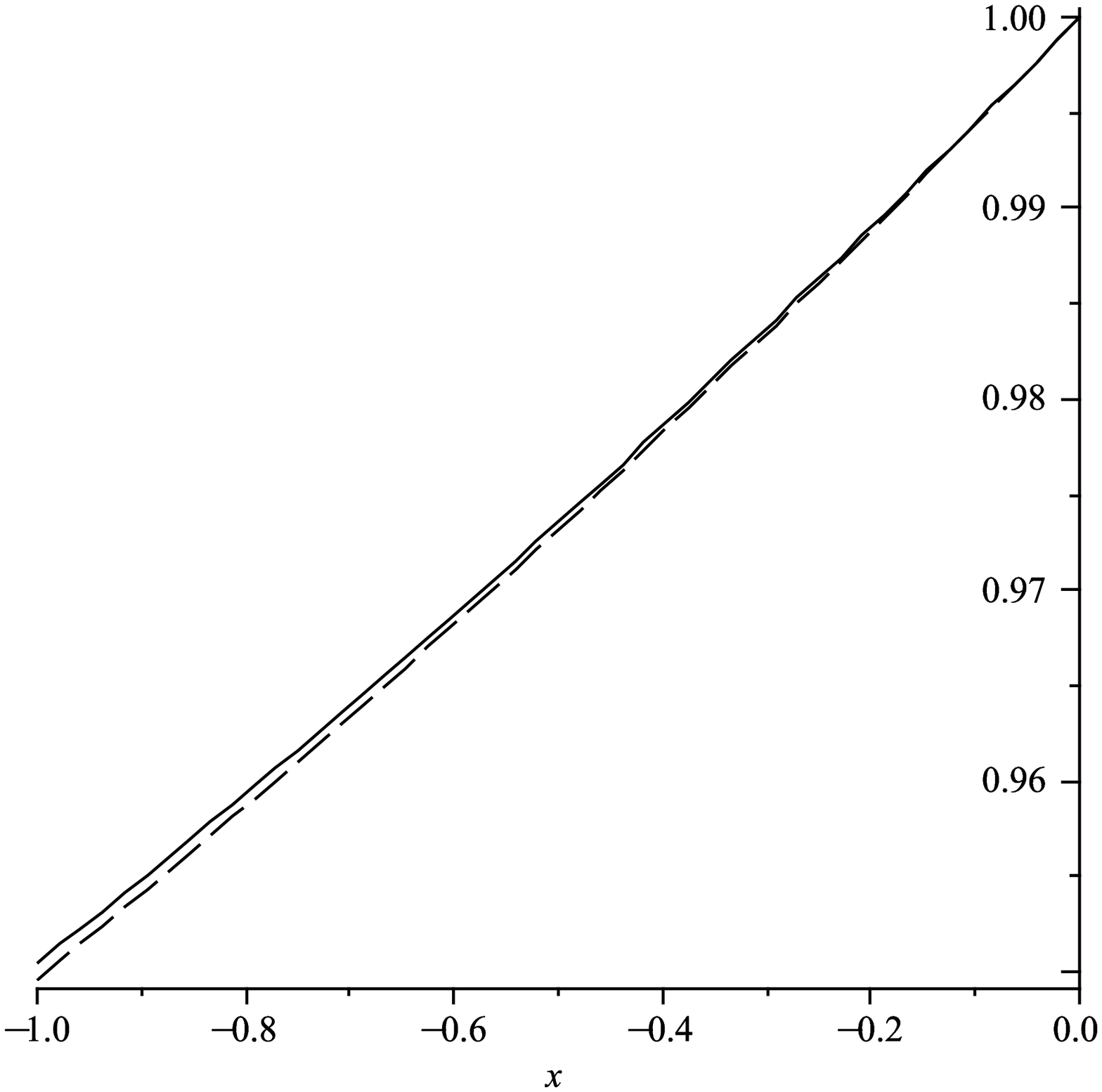} &
\includegraphics[width=4cm]{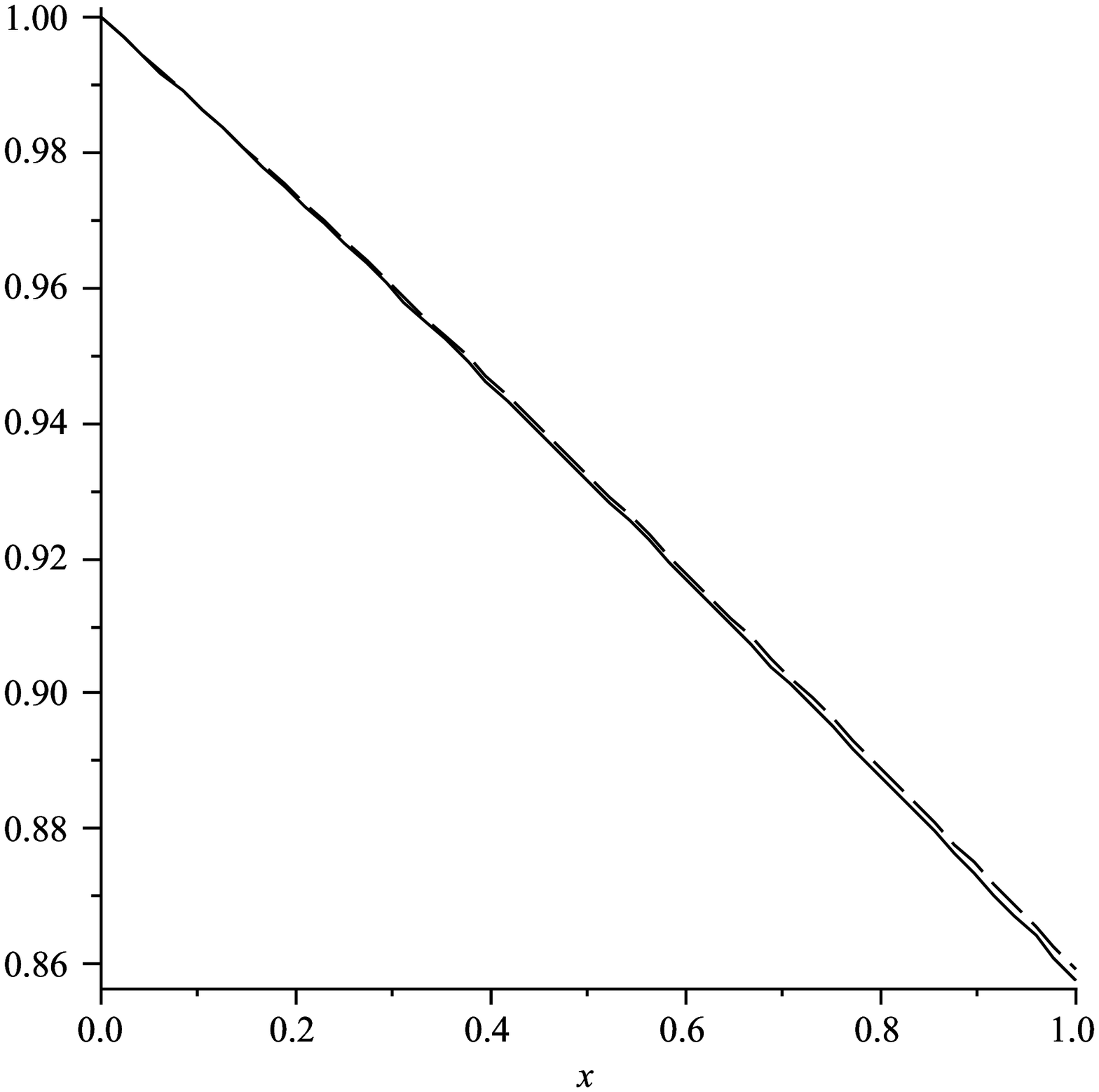}
\end{array}$
\end{center} \label{Fig1}
\caption{Comparison between the exact solutions for $\Omega_{1}(x)$ (dashed line) and
the approximate expressions  (solid line)  for a particle in an attractive delta perturbed 
by a linear potential $\lambda x$  with $\lambda=-0.1$. Left and right figures depict respectively $|\Omega_{1}(x)|$ 
on the left and right of the origin $x=0$}
\end{figure}

\begin{figure}[htbp]
\label{Fig2}
\begin{center}
\vspace{1cm}
\includegraphics[width=8cm]{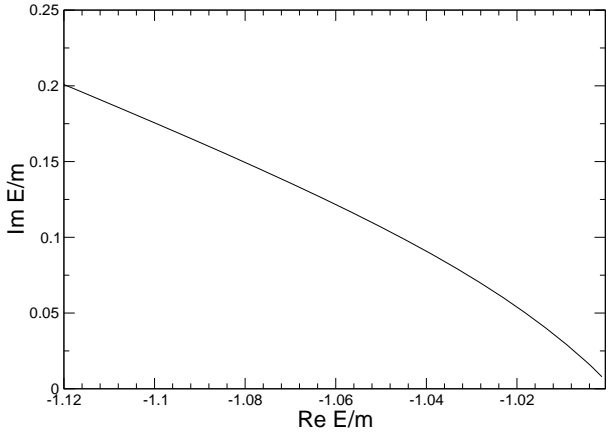}
\end{center}
\caption{ Behavior of the resonant energy  for a delta attractive interaction  with $g=-0.99 \pi$ 
perturbed by the potential 
$\Lambda \tanh(kx)$ with $\lambda=-0.1$  and $0<k<0.5$. For small values of $k$ $E$ is close to $-1$ 
and $\Im(E)$ increases as $k$ increases.}
\end{figure}

The approximation (\ref{omega1-lambda-x-aproximada}) to Eq.
(\ref{ddodirac1}) is valid when $\Lambda$ and $k$ are small with
respect to $m$. For large values of $k$ the hyperbolic potential
approaches a step potential, which is not able to sink the delta
bound energy state negative energy continuum. Fig. \ref{Fig2} shows 
the dependence of the resonant energy as the slope parameter $k$ of
the hyperbolic potential increases.

\section{Phase shifts.}

Since the perturbative potential $\Lambda\tanh(kx)$ is
asymptotically constant as  $x\rightarrow\pm\infty,$ and,
the perturbative electric field $\Lambda
k \sech(kx)^2/e$ vanishes for large values of $x$, we can use of
the scattering formalism in order to compute the phase shift
$\phi$, magnitude that will help us compute the influence of the
perturbative hyperbolic interaction on the energy spectrum of the
delta interaction. Analogous to the non-relativistic case \cite{Kleber}, 
the particle is constrained to move in a half line and  the reflection amplitude $r$ satisfies $|r|=1$. 
The phase shift $\phi$ corresponds 
to the argument of $r$ as  $\phi=-\mathrm{arg}(r)$.

\begin{figure}[htp]
\label{desf1}
\begin{center}
\includegraphics[width=8cm]{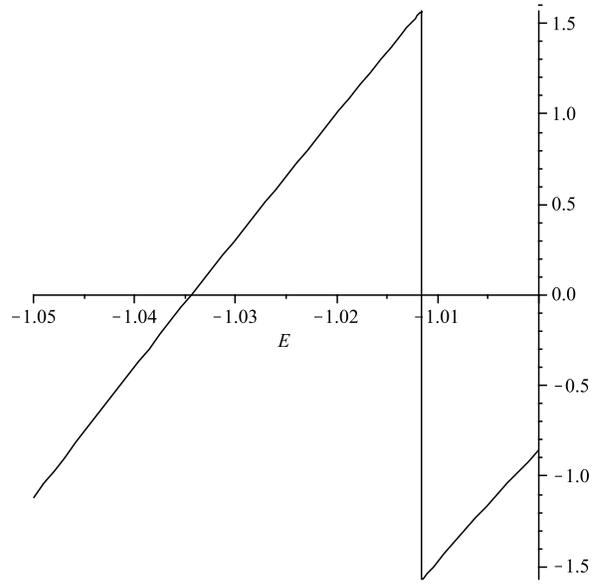}
\end{center}
\caption{Behavior of the phase shift against the energy $E$ for an attractive delta perturbed by the
potential $\Lambda \tanh(k x)$ with $\Lambda=-0.1$ and $k=0.1$. The phase shift suffers a jump of $\pi$ close
to the real part of the resonant energy $E_{res}$.}
\end{figure}

With the help of the boundary condition (\ref{dominio}) associated
with the vectorial delta interaction $-g \delta(x)$, we proceed
to calculate the reflection amplitude $r$ 
\begin{eqnarray}
tX_{1}^{t}&=&\left(X_{1}^{i}+rX_{1}^{r}\right)\cos{g}-
\left(X_{2}^{i}+rX_{2}^{r}\right)\sin{g}
\label{transmite1}\\
tX_{2}^{t}&=&-i\left(X_{1}^{i}+rX_{1}^{r}\right)\sin{g}+
\left(X_{2}^{i}+rX_{2}^{r}\right)\cos{g}, \label{transmite2}
\end{eqnarray}
where $t$  and $X^{t}$ are respectively the transmission coefficient and the transmitted wave, $X^{i}$ denotes the
incident wave.
From  Eq. (\ref{transmite1}) and  Eq. (\ref{transmite2}), we see that
\begin{equation}
\label{coeftrans}
r=\frac{X_{1}^{r}X_{2}^{i}-
X_{1}^{i}X_{2}^{r}}{\left(X_{1}^{r}X_{2}^{t}-
X_{1}^{t}X_{2}^{r}\right)\cos{g} +i\left(X_{1}^{t}X_{1}^{r}-
X_{2}^{r}X_{2}^{t}\right)\sin{g}},
\end{equation}
The solution to Eq. (\ref{ddodirac1}),  behaving as wave incoming from the left  is
\begin{equation}\label{izqinc}
\Omega_{1}^{-in}(\widetilde{z})=D_{1}(\widetilde{z}-1)^{-\frac{i\Lambda}{k}}\widetilde{z}\,^{b}{}_2F_1\left(
q,t,1+2b;\widetilde{z}\right),
\end{equation}
where
\begin{equation}
q=-a+b-\frac{i\Lambda}{k}
\end{equation}
\begin{equation}
t=-a+b-\frac{i\Lambda}{k}
\end{equation}
The component $\Omega_{2}^{-in}(\widetilde{z})$ can be obtained using Eq. (\ref{podiraca}).  
Figure \ref{desf1} shows the
behavior of the phase shift $\phi$ against the real part of the energy $E$  for $k=0.1$ and $\Lambda=-0.1$. 
It can be observed that the phase shift  increases and has
a jump of $\pi$ as it approaches the value $\pi/2$ and $E$ is close  to the real value of the
resonance $E_{res}=-1.01837+0.05058 I$  This result indicates that we are in the presence
of a bound state that dissolves itself into the negative energy
continuum.

The slope of the curve \ shown in Figure \ref{desf1} is related to
the Wigner time delay $\tau$ and can be estimated using the Heisenberg principle \cite{Newton}
\begin{equation}\label{tw2}
\Gamma\simeq\frac{\pi}{|\frac{d\phi}{dE}|_{E=E_{0}}}=\frac{1}{\tau}
\end{equation}
The Wigner time $\tau$ permits us to estimate how long the bound state
lives before it decays into the negative energy continuum.  For $\Lambda=-0.1$ and $k=0.1$
we obtain that  $\tau= 0.04622$, a value which is very close  to $\Im(E_{res})$.

The inverse of the imaginary part of the energy $E$ gives a measure
of the  mean of the-life time associated with the energy $E$
\cite{Newton,Goldberger}, a state that can be interpreted as a
positron, i.e a hole with a positive charge, created in the Dirac sea with
a finite life-time.

\section{Discussion of the results}

The vector point interaction  vectorial $-g\,\delta(x)$ described by
the boundary conditions  (\ref{dominio}) produces zero-momentum 
supercritical states for $g=\pi$ but they are not complex energy states
associated with resonant states with a Breit-Wigner profile. 

In this article we have analyzed the energy spectrum of the
one-dimensional Dirac equation in the presence of an attractive
vectorial delta point interaction when one introduces  perturbative
potentials of the form $\lambda x$ and  $\Lambda\tanh(kx)$ with small $\Lambda$ and $k$,
corresponding respectively to a constant electric field and an 
an asymptotically vanishing electric field
that reaches its maximal strength $\Lambda k$ at $x=0.$ The presence
of the perturbative electric field induces the appearance of a
resonant energy state in the one-dimensional Dirac equation. This
phenomenon is analogous to the Stark effect \cite{Galindo}.

Since the vector interaction $-g \delta(x)$ is not strong enough
to produce a real resonance\cite{FD:89,Calkin,Nogami}, we have that
the presence of the perturbative potential plays a
crucial role in the appearance of the resonant energy state. We have
studied and discussed the presence of supercriticality in two
different ways: first, studying the resonant behavior of the energy
spectrum in the vicinity of $E=-m,$ and  second, analyzing the
behaviour of the phase shift as a function of the energy.

With the help of the phase shift and the Wigner time-delay, we  have
shown that the mean life $\tau$ of the resonant bound state
decreases as the strength of the perturbative potential increases, a
result which is analogous to the one observed when a perturbative
constant electric field is considered.

The perturbative potential $\lambda x$ diverges as  $x\rightarrow\pm\infty$. This asymptotic behavior prevents us from obtaining asymptotically free solutions, therefore we cannot  use standard scattering theory for obtaining the energy resonances \cite{Goldberger}. In this sense, the  exact and approximate approaches presented in this article permits us to obtain the value of the resonances in cases where the potential does not go to zero at infinity.

\begin{acknowledgement}
This work was supported by FONACIT under project
G-2001000712.
\end{acknowledgement}

\end{document}